\documentclass[prl,showpacs,twocolumn,amsmath,amssymb,floatfix]{revtex4-1}
\usepackage[english]{babel}
\usepackage{graphicx}
\usepackage{bm}
\usepackage{color}
\usepackage{footnote}
\usepackage{hyperref}
\usepackage[applemac]{inputenc}

\begin{document}
\title{Majorana bound states from exceptional points in non-topological superconductors}  
\author{Pablo San-Jose$^1$, Jorge Cayao$^1$, Elsa Prada$^2$ and Ramón Aguado$^1$}
\affiliation{$^1$Instituto de Ciencia de Materiales de Madrid (ICMM), Consejo Superior de Investigaciones Científicas (CSIC), Cantoblanco, 28049 Madrid, Spain\\$^2$Departamento de Física de la Materia Condensada, Instituto de Ciencia de Materiales Nicolás Cabrera (INC), and Condensed Matter Physics Center (IFIMAC), Universidad Autónoma de Madrid, Cantoblanco, 28049 Madrid, Spain} 
\date{\today} 
\begin{abstract}
Recent experimental efforts towards the detection of Majorana bound states have focused on creating the conditions for topological superconductivity. Here we demonstrate an alternative route, which achieves fully localised zero-energy Majorana bound states when a topologically trivial superconductor is strongly coupled to a helical normal region. Such a junction can be experimentally realised by e.g. proximitizing a finite section of a nanowire with spin-orbit coupling, and combining electrostatic depletion and a Zeeman field to drive the non-proximitized (normal) portion into a helical phase. Majorana zero modes emerge in such an open system without fine-tuning as a result of charge-conjugation symmetry, and can be ultimately linked to the existence of `exceptional points' (EPs) in parameter space, where two quasibound Andreev levels bifurcate into two quasibound Majorana zero modes.  After the EP, one of the latter becomes non-decaying as the junction approaches perfect Andreev reflection, thus resulting in a Majorana dark state (MDS) localised at the NS junction. We show that MDSs exhibit the full range of properties associated to conventional closed-system Majorana bound states (zero-energy, self-conjugation, $4\pi$-Josephson effect and non-Abelian braiding statistics), while not requiring topological superconductivity.
\end{abstract}
\maketitle
 
\section{{Introduction}}
The emergence of topologically protected Majorana zero modes in topological superconductors has recently entered the spotlight of condensed matter research   \cite{Alicea:RPP12,Leijnse:SSAT12,Beenakker:ARCMP13,Stanescu:JOPCM13,Elliott:RMP15}
One of the main reasons is the remarkable prediction that such Majorana bound states (MBSs), also known as Majorana zero modes, should obey non-Abelian braiding statistics \cite{Kitaev:PU01,Ivanov:PRL01}, much like the 5/2 states in the fractional Hall effect, without requiring many-body correlations. 
It has been argued that the successful generation, detection and manipulation of MBSs would open the possibility of practical topologically protected quantum computation \cite{Nayak:RMP08,Sarma:15}. 
Despite impressive experimental progress  \cite{Mourik:S12,Deng:NL12,Das:NP12,Rokhinson:NP12,Finck:PRL13,Churchill:PRB13,Lee:NN14,Hart:NP14,Pribiag:NN15,Nadj-Perge:S14}, such ambitious goals have still not been conclusively achieved.

A number of practical proposals have been put forward aiming to generate the conditions for the spontaneous emergence of robust MBSs in real devices. Some of the most studied ones are based on proximising topological insulators \cite{Fu:PRL08} or semiconductor nanowires \cite{Lutchyn:PRL10,Oreg:PRL10}. The core challenge in all these proposals has been to artificially synthesise a topologically non-trivial superconductor with a well-defined and robust topological gap\cite{Takei:PRL13}. The bulk-boundary correspondence principle dictates that the superconductor surface is then host to topologically protected MBSs. Creating a topological gap is arguably the main practical difficulty of such proposals, particularly since topological superconductors are rather sensitive to disorder.

\begin{figure}
   \centering 
   \includegraphics[width=0.98\columnwidth]{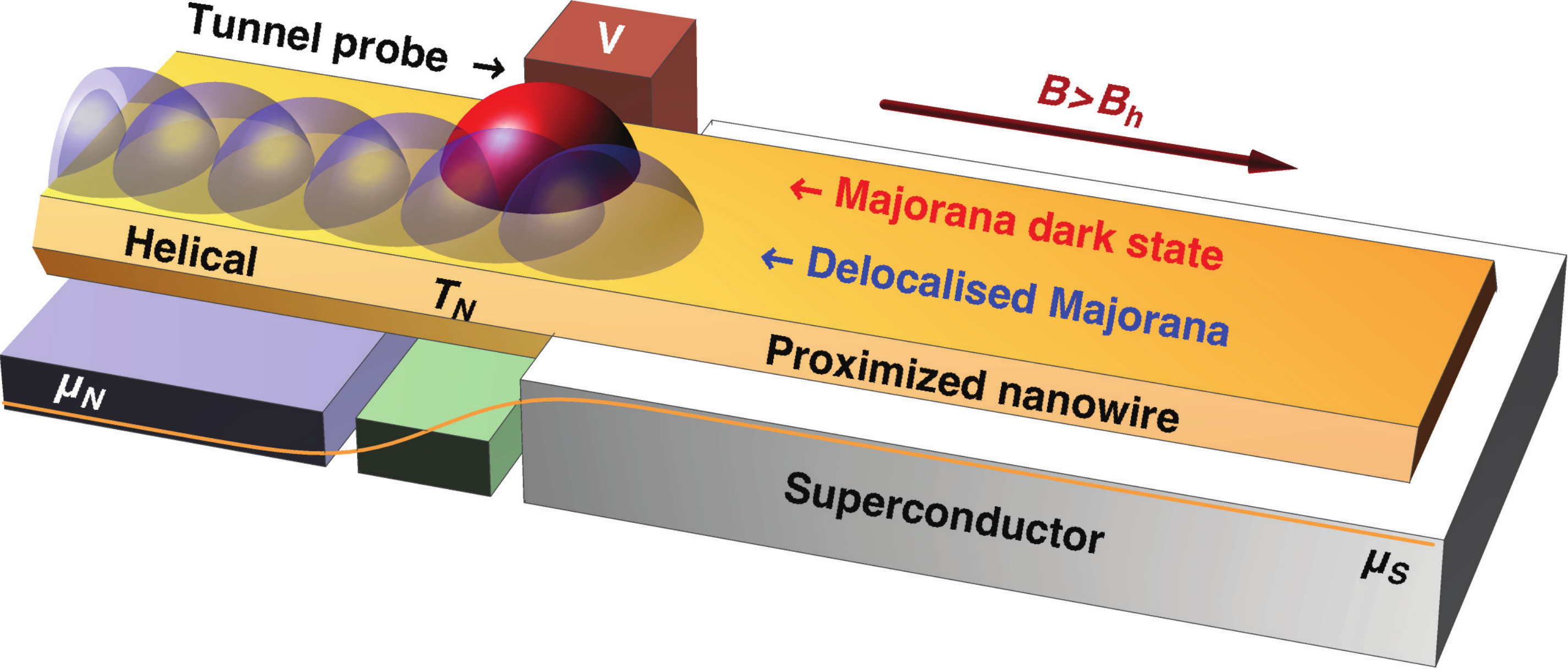}
   \caption{{\bf A sketch of a semiconductor nanowire, partially proximitized with a conventional superconductor on the right side}. The normal side may be depleted (small Fermi energy $\mu_N$), and may become helical under a Zeeman field, $B>B_h\equiv\mu_N$, while the superconducting side remains topologically trivial at small fields. For sufficiently transparent junctions in the Andreev limit ($\Delta\ll\mu_S$), this results in Majorana dark state bound to the junction, in red.} 
   \label{fig:WireSketch} 
\end{figure}

In this work we demonstrate an alternative scheme for the creation of MBSs that does not require topological superconductivity at all.  The possibility of engineering Majoranas in topologically trivial setups has been studied in other contexts before. It has been shown, for example, that topological excitations, and MBSs in particular, may arise in trivial superconductors under adequate external driving \cite{Jiang:PRL11,Kundu:PRL13}, similarly to the mechanism behind Floquet topological insulators \cite{Lindner:NP11}. Also, cold-atom systems with specifically engineered dissipation \cite{Diehl:NP11,Bardyn:PRL12} may relax into a topologically non-trivial steady state that are host to dark states at zero energy with Majorana properties.  
Our approach is implemented in a solid state setup and is based on proximitized semiconductor nanowires. In its topologically trivial regime, such a wire will not generate MBSs when terminated with vacuum (i.e. at a closed boundary). Its spectrum is instead a set of Bogoliubov quasiparticles that can be seen as pairs of Majoranas hybridized to finite energy. By creating a sufficiently transparent normal-superconductor (NS) junction at one end of the wire, we create a different kind of \emph{open} boundary, to which the bulk-boundary correspondence principle does not apply. Such a high-transparency junction can be fabricated by proximitizing only one half of a pristine semiconducting nanowire (Fig. \ref{fig:WireSketch}).
We demonstrate that, as one tunes the normal side into a helical (half-metallic) regime via a parallel Zeeman field, one Majorana pair becomes decoupled into two zero energy resonances. One of which is subsequently removed into the reservoir, leaving behind a stable Majorana `dark state' (MDS) at the NS junction without requiring a non-trivial superconductivity. (A dark state here is defined as a bound state that despite having an energy embedded in a continuum of delocalized excitations is orthogonal to them and, therefore, non-decaying.)

The emergence of these MDSs cannot be described using the conventional band topology language, but rather needs to be understood in the context of open quantum systems. Unlike in closed systems, eigenstates in open systems decay with time, as the state leaks into the reservoir. Hence, their energies $\epsilon_p=E_p-i\Gamma_p$ are no longer real, but have a negative imaginary part that represents this decay rate $\Gamma_p$. Such a complex spectrum is sometimes modelled by a non-Hermitian Hamiltonian. A more precise and general description is  obtained by considering the analytic continuation of the scattering matrix $S(\omega)$, where the energy $\omega$ of incoming states from the reservoir  is allowed to extend into the lower complex half-plane. The analogous to the real eigenvalues of the closed system then becomes the poles of $S(\omega)$ for the open system. In NS junctions, all such poles come in pairs $\pm E_p-i\Gamma_p$ for all poles with non-zero real part $E_p$, due to the charge-conjugation symmetry of the Nambu representation \cite{Pikulin:JL12,Pikulin:PRB13}. In this sense, poles with zero real part $E_p=0$  are special, as they do not come in pairs. Their total number $Z$ (which excludes any pole with zero real \emph{and} imaginary part) has a very important meaning in open NS junctions, and defines the analogue of band topology of a closed quantum system. Indeed, it has been shown that the topology of the scattering matrix in quasi-1D NS junctions is classified by the invariant $\nu=Z\,\mathrm{mod}\,2$,  i.e. the parity of the number of poles with zero real energy, with $\nu=1$ signalling an  open system with non-trivial topology from the point of view of scattering\cite{Pikulin:JL12,Pikulin:PRB13}.

In terms of its S-matrix poles, the topological transitions of an NS junction follow a characteristic pattern. Consider a trivial S-matrix ($\nu=0$). As a given parameter of the system is varied, a pair of poles $\epsilon_p=\pm E_p-i\Gamma_p$ first approach each other and become degenerate at the imaginary axis, $E_p=0$. This degeneracy is the open-system counterpart of a band inversion in a closed system, and is known as an exceptional point (EP) \cite{Kato:95,Moiseyev:11,Berry:CJOP04,Heiss:JPA12,Heiss:PRE00}. It differs from a closed-system degeneracy in that the corresponding eigenstates do not remain orthogonal, but rather coalesce into one. {Exceptional points have been extensively studied in photonics  where they have been shown to give rise to novel phenomena unique to open systems \cite{Bender:PRL98,Klaiman:PRL08,Regensburger:N12,Guo:PRL09,Liertzer:PRL12,Brandstetter:NC14,Peng:S14,Poli:NC15}. Their implications in electronic systems, however, have been seldom discussed \cite{Mandal:EEL15,Malzard:15}.}

After the exceptional point, the two degenerate poles branch along the imaginary axis, and their decay rates bifurcate into different values $\Gamma_0<\Gamma_1$ (see Fig. \ref{fig:OpenKitaev}d for an example). The exceptional point thus involves a change of $Z$ by 2, but $\nu=0$ remains unchanged. If $\Gamma_0$ evolves towards zero (or close enough to zero for all practical purposes), it is said that the corresponding pole is \emph{buried}, and it is excluded from the $N$ count, effectively signalling a change of topology $\nu=1$.  Crucially, the existence of a buried pole implies the existence of a zero-energy non-decaying (dark) state somewhere in the system with Majorana properties. In this sense, S-matrix topology is a true generalization of the band-structure topology of closed systems, and has the same implications in terms of topologically protected excitations, albeit in the context of open systems. It is also closely linked to the existence of an exceptional point in the system that occurs before the pole burying, in the trivial $\nu=0$ phase.

Here we show that, while at weak couplings between the normal environment and the superconductor a non-trivial S-matrix implies that the superconductor is also non-trivial in isolation, this is not the case at strong couplings. In specific but experimentally relevant conditions, a sufficiently transparent junction between a normal metal and a \emph{trivial} superconductor has a non-trivial S-matrix with $\nu=1$, and is thus host to a Majorana dark state. The required conditions are: (1) the system should have a finite spin-orbit coupling at the contact, (2) the normal part of the junction should be sufficiently depleted (small Fermi energy $\mu_N$) and polarised by a Zeeman field $B$ into a helical half-metallic phase $B>\mu_N$, (3) the normal transmission of the junction should be close to one, and (4) the trivial superconductor should be in the Andreev limit $\Delta\ll \mu_S$, where $\Delta$ is the superconducting gap and $\mu_S$ is its Fermi energy. The rationale of these conditions is to achieve good Andreev reflection of helical carriers from the normal side, which generates a MDS strongly localized at the junction. The intuitive mechanism behind the process is as follows. When isolated, the trivial superconducting wire is host to a Majorana \emph{pair} at each end, which is strongly hybridized into a fermionic Bogoliubov state, with real energies $\pm E_p$. As the contact is opened onto a helical wire (which has a single decay channel), one (and only one) of the two Majoranas escapes into the reservoir (blue state in Fig. \ref{fig:WireSketch}), leaving behind the orthogonal Majorana as a dark state (red), pinned at zero energy since it no longer overlaps with the escaped Majorana. This process takes place formally by crossing an exceptional point bifurcation of the $\pm E_p$ poles into poles $i\Gamma_{0,1}$ on the imaginary axis.
As the conditions above are fulfilled, the zero energy dark state becomes truly non-decaying $\Gamma_0\to 0$. Deviations from these conditions result in a residual decay rate $\Gamma_0$. The dark state is then a sharp Majorana resonance centered at zero, but with Majorana properties surviving at times shorter than $\tau_0=1/\Gamma_0$.

We analyse the generation of a MDSs in two different models. First we consider the problem of the open multimode Kitaev wire, and see how the above scenario plays out in this simple model. Then we consider a more realistic model of an NS contact in a proximitized semiconducting wire. We describe the experimental signatures associated to the MDS. We also analyze its properties of the dark state to demonstrate it indeed shares all the characteristics of a Majorana bound state from a closed topological system, including self-conjugation, locality and neutrality, and the appearance of uniform charge oscillations, 4$\pi$ Josephson effect and non-Abelian braiding when combined with another MDS. We show that the residual decay rate of the MDS in non-ideal conditions depend exponentially with junction lengthscales, just like the Majorana splitting in closed topological superconductors.

\section{Results}

\subsection*{Exceptional points and Majorana dark states in the open Kitaev model} 
\begin{figure} 
   \centering
   \includegraphics[width=0.98\columnwidth]{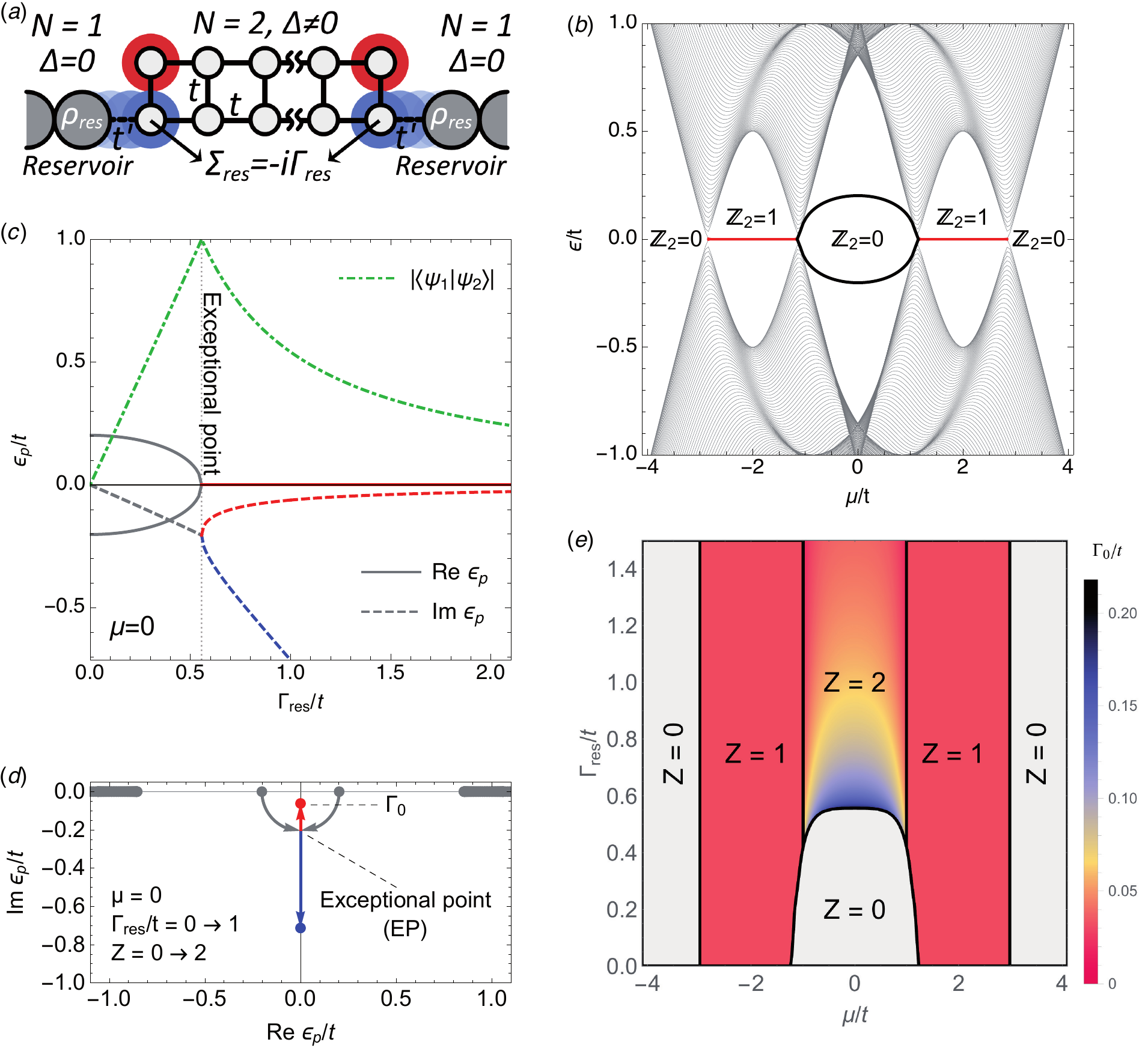}
   \caption{{\bf Exceptional points in the open two-mode Kitaev model}. (a) Sketch of the system. (b) Bandstructure in the closed limit $\Gamma_\textrm{res}=0$ for $\Delta=0.5 t$ and $L=140$, exhibiting trivial ($\nu=0$) and non-trivial ($\nu=1$) phases as a function of chemical potential $\mu$. (c,d) Evolution of the lowest lying complex eigenvalues in the trivial phase at $\mu=0$ [thick black in (b)], as the coupling to the reservoirs $\Gamma_\textrm{res}$ is increased. An exceptional point is crossed at a critical $\Gamma_\textrm{res}\approx 0.55 t$, which results in the emergence of $Z=2$ eigengalues per edge with $\mathrm{Re}\,\epsilon_p=0$. At the exceptional point both the eigenvalues and the eigenstates coalesce [dot-dashed green curve in (c)]. (e) Full phase diagram for $Z$ as a function of $\mu$ and $\Gamma_\textrm{res}$. Colors indicate the decay rate $\Gamma_0$ of the most stable of the $Z$ eigenvalues, red in panel (d).}
   \label{fig:OpenKitaev}  
\end{figure}

We first study the emergence of stable zero-energy dark states in an open toy model, in preparation for the more realistic calculation presented later for proximitized semiconducting wires. We consider a natural multimode D-class extension \cite{Potter:PRL10, Wakatsuki:PRB14} of the original Kitaev model \cite{Kitaev:PU01}. It is a finite length (i.e. closed) quasi-1D chain of spinless fermions with a $p_x+i p_y$ superconducting pairing 
with interesting topological properties. It may be written as a nearest-neighbour tight-binding Hamiltonian in an $L\times N$ square lattice,
\begin{eqnarray} \label{Kitaev} 
H_\textrm{closed}&=&-\mu\sum_{\vec n}c_{\vec n}^\dagger c^{\phantom{\dagger}}_{\vec n}-\sum_{\langle\vec n, \vec n'\rangle}t c_{\vec n'}^\dagger c^{\phantom{\dagger}}_{\vec n} \nonumber\\
&&-\sum_{\langle\vec n, \vec n'\rangle}\Delta^{\phantom{\dagger}}_{\vec n'-\vec n} c_{\vec n'}^\dagger c^\dagger_{\vec n}+\mathrm{H.c}
\end{eqnarray}
The hopping amplitude is $t$, $\vec n=(n_x,n_y)$ are integer site indices, and $\mu$ is the chemical potential. The $p_x+i p_y$ pairing is implemented by $\Delta_{\pm\hat{x}}=\pm\Delta$ and $\Delta_{\pm\hat{y}}=\pm i\Delta$. For $N>1$, $t>\Delta$  and finite length $L\gg 1$ (wire terminated by vacuum), this model exhibits an even-odd effect in its spectrum \cite{Potter:PRL10} as $\mu$ is varied, see Fig. \ref{fig:OpenKitaev}b. Odd (non-trivial) phases develop a single zero-energy MBS at each end of the wire (in red, Majorana number $\nu=1$), while even (trivial) phases have none ($\nu=0$). These MBSs arise as a result of the bulk-boundary correspondence principle when the bulk bandstructure of the wire becomes topologically non-trivial, with $\nu$ as the topological invariant\cite{Altland:PRB97,Qi:RMP11,Wakatsuki:PRB14}.

We now analyse the behaviour of the spectrum with a different type of boundary. Instead of vacuum we couple the wire to a normal single-mode reservoir at each end. This is a very different boundary (it is gapless), to which the bulk-boundary correspondence principle is not applicable. We will show how a strong enough coupling to this specific single-mode reservoir may allow certain topologically trivial phases ($\nu=0$, no MBSs with vacuum termination) to develop stable bound states at the contacts, or dark states 
pinned to zero energy, and closely related to the MBSs of non-trivial phases. 

For concreteness we consider the $N=2$ Kitaev wire, which is the simplest that manifests this phenomenon. The wire is coupled to single-mode ($N=1$) normal reservoirs at each end, as shown in Fig. \ref{fig:OpenKitaev}a. Its effect is modelled through a non-Hermitian self-energy $\Sigma_\textrm{res}$ on one end site,
which results in a non-Hermitian effective Hamiltonian for the open multimode Kitaev wire
\begin{eqnarray} \label{Heff}
H^\textrm{eff}_\textrm{open}&=&H_\textrm{closed}+\Sigma_\textrm{res}\\
\Sigma_\textrm{res}&\approx& -i \Gamma_\textrm{res} \left(c_{(0,0)}^\dagger c^{\phantom{\dagger}}_{(0,0)}+c_{(L,0)}^\dagger c^{\phantom{\dagger}}_{(L,0)}\right)\nonumber
\end{eqnarray}
The local decay rate $\Gamma_\textrm{res}$ depends on the (constant) local density of states of the reservoir at the contact $\rho_\textrm{res}$, and the hopping amplitude $t'$ into the reservoir, $\Gamma_\textrm{res}=\pi |t'|^2\rho_\textrm{res}$. The shift $\mathrm{Re}\,\Sigma_\mathrm{res}$ is neglected for the moment. 

The eigenvalues of $H^\textrm{eff}_\textrm{open}$ are complex $\epsilon_p=E_p-i\Gamma_p$, and represent the real energy $E_p$ and decay rate $\Gamma_p$ into the reservoirs of quasibound states in the wire. They are also the poles of the scattering matrix $S(\omega)$ from the reservoir for complex $\omega$. Causality implies $\mathrm{Im}\,\epsilon_p<0$, while the charge-conjugation symmetry of the Bogoliubov-de Gennes description guarantees that if $\epsilon_p$ is an eigenvalue, then $-\epsilon_p^*$ is also. This symmetry classifies eigenvalues in conjugate pairs, unless they lie exactly on the imaginary axis, in which case they may have no partner. The number of such lone, purely imaginary (or zero) eigenvalues per edge is denoted by $Z$, and carries a profound physical significance related to the topology of the scattering matrix at that contact. The number $Z$ allowed Pikulin and Nazarov \cite{Pikulin:JL12,Pikulin:PRB13} to classify the scattering matrices $S(\omega)$ of generic NS junctions into trivial (even $Z$) and non-trivial (odd $Z$) topological classes, and connect it to bulk topological invariants by $\nu=Z\,\mathrm{mod}\, 2$. $Z$ is moreover a robust quantity. An unpaired quasibound state with $\epsilon_p=-i\Gamma_p$ cannot acquire a finite real energy $E_p$ through any small perturbation, just change its decay rate $\Gamma_p$. This robust pinning to zero real energy, protected by charge-conjugation symmetry, is reminiscent of the topologically protected zero energy pinning of MBSs in closed topological superconductors.

The evolution of the complex $\epsilon_p$ as a function of $\Gamma_\mathrm{res}$ for $\mu=0$ is shown in Fig. \ref{fig:OpenKitaev}(c,d), and the phase diagram of the model in the full $(\mu,\Gamma_\mathrm{res})$ plane is shown in Fig. \ref{fig:OpenKitaev}e. The trivial $\nu=Z=0$ phase at $\Gamma_\mathrm{res}=0$ around $\mu=0$ is characterized  by two subgap non-zero eigenvalues resulting from the hybridization of two MBSs at each edge [black curves in panel (b)]. As $\Gamma_\mathrm{res}$ is increased, this trivial phase experiences a transition into $Z=2$ through the fusion of these two levels. This is an instance of an `exceptional point', a generic feature of open systems, or in general of non-Hermitian matrices with certain symmetries \cite{Heiss:JPA12,Malzard:15}, at which two complex eigenvalues become degenerate and bifurcate in the complex plane. One such bifurcation is shown in panels (c,d). At the exceptional point (here $\Gamma_\mathrm{res}\approx 0.55 t$ for $\mu=0$), the corresponding eigenstates coalesce [their overlap reaches one, dot-dashed green curve in panel (c)], in contrast to the Hermitian case where degenerate eigenstates remain orthogonal. 
 
After crossing the exceptional point, the two hybridized Majoranas become exact zero modes, albeit with different lifetimes, without any fine tuning. One of the two eigenvalues at each contact gradually approaches the origin [red in panel (d)], i.e. its imaginary part also approaches zero. We denote this imaginary part, or decay rate, by $\Gamma_0$. Note that $\Gamma_0$ is not of order $\Gamma_\mathrm{res}$. In fact $\Gamma_0$ is \emph{suppressed} as $\Gamma_\mathrm{res}$ increases, see panels (c,e). As a result, the corresponding state evolves into a long-lived zero-energy resonance localised at an increasingly transparent contact. The suppression of its decay $\Gamma_0$ is a result of the specific single-mode nature of the reservoirs considered here, and is not related to any change in the bulk topology, which remains trivial $\nu=0$. 

\begin{figure}
   \centering
   \includegraphics[width=\columnwidth]{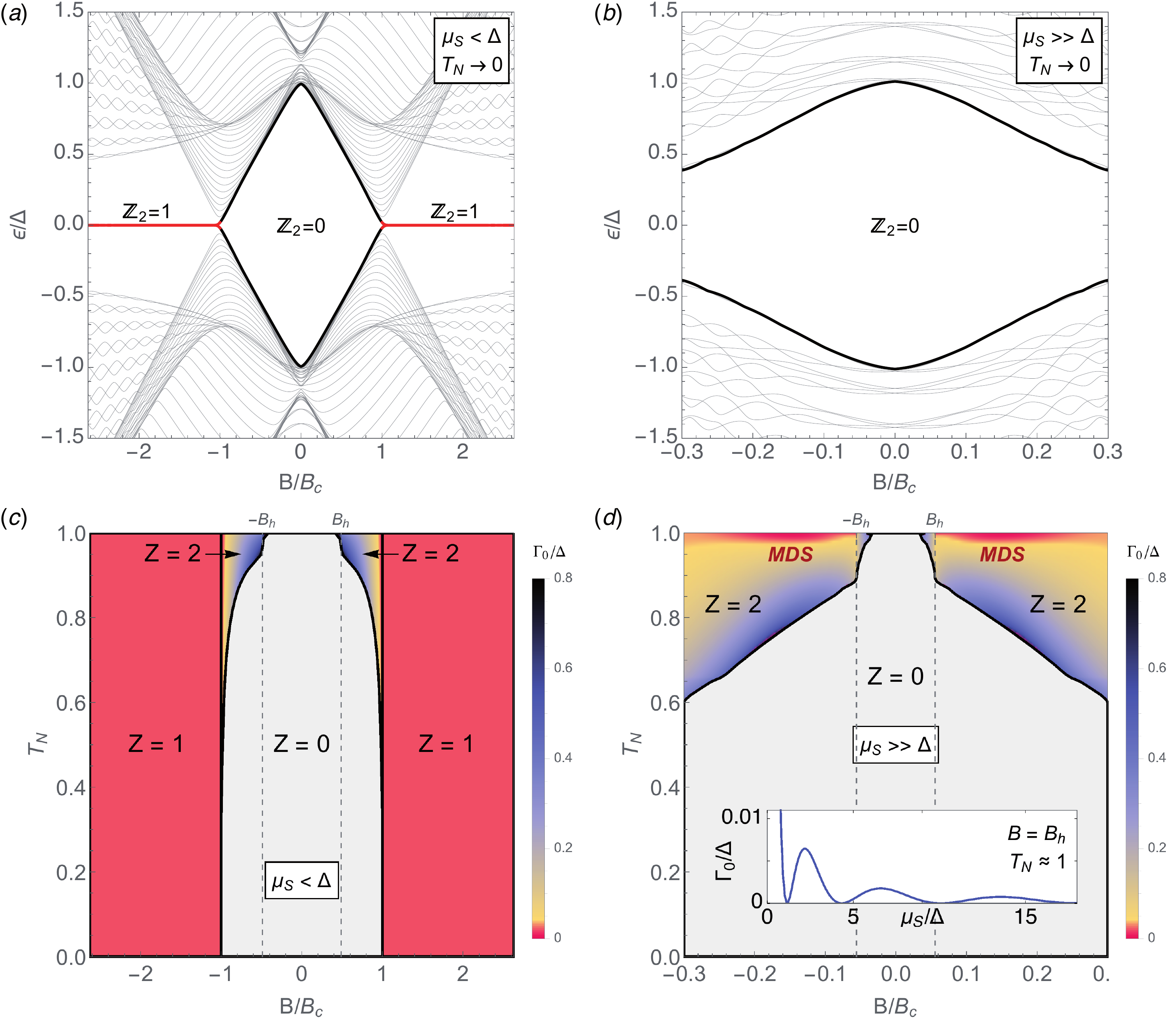}
   \caption{{\bf Exceptional points in a proximitized Rashba nanowire}. (a,b) The spectrum for an InSb proximitized wire ($\Delta=0.25$ meV, $\alpha_\mathrm{SO}=20$ meV nm, $m^*=0.015 m_e$, $g=40$) at vanishing NS transparency both for $\mu_S/\Delta\approx 0.5$ (a) and $\mu_S/\Delta\approx 10$ (b), for the same range of  Zeeman field $B$. (c,d) The corresponding phase diagram of the NS junction in the $B-T_N$ plane. Colors denote the residual decay rate $\Gamma_0$ after the exceptional point (thick black line). Panel (d) in the Andreev limit shows the formation of Majorana dark states (MDSs) in the trivial phase $B<B_c$ at high transparency $T_N$ [red regions]. Inset: the residual decay rate of the MDSs vanishes as $\mu_S$ is pushed into the Andreev limit $\mu_S\gg \Delta$. For the realistic parameters of the simulations, this residual decay rate corresponds to a lifetime of the MDSs of $\sim$0.2 microseconds.}
   \label{fig:OpenWire} 
\end{figure}

Physically, the above bifurcation can be understood as follows. In the $N=2$ trivial phase around $\mu=0$, the Kitaev wire actually hosts two Majorana bound states at each end, initially at zero energy, that owing to their strong overlap hybridize into finite energy (thick black line in Fig. \ref{fig:OpenKitaev}b). These two states are schematically represented by the red and blue circles in Fig. \ref{fig:OpenKitaev}a. As the wire is opened to the single-mode reservoir, only one combination of these two states (blue circle) can scape across the transparent junction into the reservoir while its overlap with the orthogonal combination (red circle), which remains localized, is suppressed. As a result, the decay rates of the two states bifurcate at the exceptional point while their real part becomes \emph{exactly} zero. Thus, the fact that the reservoir has a single mode into which only one of the two Majoranas may decay is essential in order to yield one decoupled dark state at a transparent junction. Here, transparent should be understood as perfect Andreev reflection probability. This implies perfect delocalisation of one Majorana, and a decoupled zero-energy dark state.
To have a perfect Andreev reflection, however, it is not enough to have a transparent contact in the normal phase ($\Delta=0$). It is also necessary that the pairing is a small perturbation to the normal phase ($\Delta\ll t$, Andreev limit). Otherwise a non-zero normal reflection will arise at the NS junction, leading to a finite residual decay $\Gamma_0$. As we approach the Andreev limit, the suppression of the residual decay rate $\Gamma_0$ at large $\Gamma_\mathrm{res}$ is very fast, decreasing exponentially with the ratio $t/\Delta$ (not shown). Hence, the Majorana resonance quickly becomes a proper MBS for $t,\Gamma_\mathrm{res}\gg\Delta$.

It has been shown that realistic models for an isolated, Zeeman-polarised, proximitized 1D Rashba wire belongs to the same topological class as the Kitaev model \cite{Alicea:RPP12}. Such models also develop a $\nu=1$ topological phase with MBSs for strong enough Zeeman fields, while they remain trivial $\nu=0$ at low fields. In the following we study the emergence at exceptional points of zero energy dark states with Majorana properties in open topologically trivial Rashba wires.

\subsection*{Majorana dark states in a proximitized Rashba wire}
In recent years, experimental progress has been reported towards the detection of MBS in Rashba nanowires  \cite{Mourik:S12,Deng:NL12,Das:NP12,Rokhinson:NP12,Finck:PRL13,Churchill:PRB13,Lee:NN14}. These efforts were in large part stimulated by the prediction by Lutchyn \textit{et al.}\cite{Lutchyn:PRL10} and Oreg \textit{et al.} \cite{Oreg:PRL10} that these type of systems would undergo a topological transition into an effective p-wave superconducting phase when a Zeeman field $B$ parallel to the wire exceeds a critical value $B_c$. We now consider models relevant to single-mode Rashba wires, and demonstrate the formation of MDSs for $B<B_c$.

Following Refs. \cite{Lutchyn:PRL10,Oreg:PRL10}, we model a thin proximitized Rashba nanowire under a Zeeman field by a spinful 1D tight-binding chain, 
\begin{eqnarray} \label{model}
H_\textrm{S}&=&(2t-\mu_S)\sum_{\sigma n}c_{\sigma n}^\dagger c^{\phantom{\dagger}}_{\sigma n}+\sum_{\sigma n}\Delta\, c_{\sigma n}^\dagger c^\dagger_{\bar\sigma n}+\mathrm{H.c} \\
&&-\sum_{\sigma, \langle n,n'\rangle}t\, c_{\sigma n'}^\dagger c^{\phantom{\dagger}}_{\sigma n} -i\sum_{\sigma,\sigma' \langle n,n'\rangle}t^{\mathrm{SO}\phantom{\dagger}}_{n'-n}c_{\sigma' n'}^\dagger \sigma^{y}_{\sigma'\sigma}c^{\phantom{\dagger}}_{\sigma n}\nonumber\\
&&+\sum_{\sigma,\sigma' n}B\,c_{\sigma' n}^\dagger \sigma^{x}_{\sigma'\sigma}c^{\phantom{\dagger}}_{\sigma n}\nonumber
\end{eqnarray} 
The parameters of the model are the chemical potential measured from depletion $\mu_S$, the hopping $t=\hbar^2/(2m^*a_0^2)$, where $m^*$ is the effective mass and $a_0$ is the lattice spacing, the induced pairing $\Delta$, the SO hopping $t^\mathrm{SO}_{\pm 1}=\pm \frac{1}{2}\alpha_\mathrm{SO}/a_0$, where $\alpha_\mathrm{SO}=\hbar^2/(m^*\lambda_\mathrm{SO})$ is the SO coupling and $\lambda_\mathrm{SO}$ is the SO length, and the Zeeman field $B=\frac{1}{2}g\mu_B \mathcal{B}_x$, where $g$ is the g-factor, and $\mathcal{B}_x$ is the magnetic field along the wire. In what follows we present simulations with parameters corresponding to an InSb proximitized wire \cite{Mourik:S12} ($\Delta=0.25$ meV, $\alpha_\mathrm{SO}=20$ meV nm, $m^*=0.015 m_e$, $g=40$). Like for the Kitaev wire, we will consider both an isolated proximitized wire of finite length, and an open NS contact between proximitized and non-proximitized sections of a nanowire, Fig. \ref{fig:WireSketch}. The latter is assumed infinite (see the Appendix for finite length effects), and is modelled by the same Hamiltonian, albeit with $\Delta=0$ and a $\mu_N$ in place of $\mu_S$. The normal-state average transparency per mode $T_N$ of the contact is physically controlled by a electrostatic gating in an actual device, and is modelled here either by a hopping $t'\leq t$ across the contact, or by a spatial interpolation between $\mu_N$ and $\mu_S$ {across a certain contact length $L_C$ determined by the distance of the wire to the depletion gate ($\Delta$ is always abrupt, see Appendix). Note that if the density of defects in the wire is small, $T_N\sim 1-\exp(-L_C/\lambda)$, with $\lambda$ a lengthscale of the order of the average Fermi wavelength.}
 
Topologically, the isolated nanowire belongs, for finite $B$, to the same one-dimensional D-class as the multimode Kitaev wire of the preceding section. For $|B|$ smaller than a critical $B_c=(\mu_S^2+\Delta^2)^{1/2}$, the nanowire is trivial ($\nu=Z=0$). 
As $|B|$ exceeds $B_c$, the topological invariant becomes non-trivial ($\nu=1$) through a band inversion, see Fig. \ref{fig:OpenWire}a, much like the transition at $|\mu|\approx t$ in Fig. \ref{fig:OpenKitaev}b, with the peculiarity that the two hybridized Majoranas in the trivial $|B|<B_c$ phase are not deep inside the gap, but at the band edge. As $\mu_S$ grows, $B_c$ quickly becomes unrealistically large, and the nanowire remains trivial for all reasonable fields, Fig. \ref{fig:OpenWire}b. {(A large $\mu_S$, incidentally, is the natural experimental regime, since the superconductor will typically transfer charge to the proximitized section of the wire that is difficult to deplete due to screening.)}

As the nanowire contains a non-proximitized normal section, Fig. \ref{fig:WireSketch}, a phenomenology similar to the open $N=2$ Kitaev wire arises. We define the effective Hamiltonian like in Eq. (\ref{Heff}), albeit with the exact self energy from the normal portion of the wire evaluated at zero frequency $\Sigma_N(\omega=0)$ (see Appendix). As the coupling to the reservoirs increases (the junction transparency $T_N$ grows), two eigenvalues drop out from the band edge into the lower complex plane, and merge at the imaginary axis at an exceptional point, much like in Fig. \ref{fig:OpenKitaev}d.
In contrast to the Kitaev model, however, this exceptional point is only reached if the Zeeman field exceeds a certain value $|B|\gtrsim B_h\equiv \mu_N$, see dashed lines in \ref{fig:OpenWire}(c,d). This $B_h$ is the field required for the normal nanowire to become \emph{helical}. For $|B|>B_h$, the normal nanowire hosts a single propagating mode, with the other spin sector completely depleted by the Zeeman field, and behaves as a single-mode reservoir like the one discussed for the Kitaev wire. Note that this helical regime should be achievable using electrostatic gating, since it only requires a sufficient depletion of the non-proximitized section of the semiconductor nanowire, {unscreened by the superconductor}.

After crossing the exceptional point (region above thick black curves in \ref{fig:OpenWire}[c,d]), the scattering matrix $S(\omega)$ at the trivial NS contact acquires $Z=2$ purely imaginary poles. One of the two moves towards the origin. The asymptotic decay rate $\Gamma_0$ in the limit $T_N\to 1$ is not vanishing in general, so that the corresponding states should be denoted as Majorana resonances \cite{Sau:PRL12}. However, in the experimentally relevant Andreev limit $\mu_S\gg\Delta$, Fig. \ref{fig:OpenWire}d, the asymptotic $\Gamma_0$ vanishes exponentially with $\mu_S/\Delta$ (see inset). When the wire is tuned into the regime $\mu_S\gg\Delta$ and the contact is made sufficiently transparent, the Majorana resonances are stabilised into proper non-decaying MDSs. For example, for the realistic parameters of the simulations in Fig. \ref{fig:OpenWire}d the lifetime for the minimum widths (see inset) corresponds to $\sim$0.2 microseconds. While this is already a rather long time, this is not an upper bound since even longer lifetimes can in principle be obtained by increasing $\mu_S$.

The interpretation of this mechanism is the same as in the Kitaev model. While two MBSs at opposite ends of an isolated topological nanowire can be considered exact zero modes up to exponentially small corrections (in the wire length) coming from their mutual overlap, an MDSs (red in Fig. \ref{fig:WireSketch}) also becomes an exact Majorana zero mode without any fine tuning by a suppression of its overlap with its sibling Majorana (blue in Fig. \ref{fig:WireSketch}), which escapes into the helical reservoir. It is important to note that, in contrast to isolated topological wires, any residual overlap that remains after the exceptional point does not translate into a finite energy splitting, but rather into a residual decay rate $\Gamma_0$.

\subsection*{Physical properties of Majorana dark states}

Having established the emergence of zero energy dark states at a transparent helical metal-trivial superconductor junction in the Andreev limit, we now turn to the analysis of the physical properties of said states, and compare them to conventional MBSs. We will  study their signatures in transport, their wavefunction locality, particle-hole conjugation, their charge neutrality, uniform charge oscillations and finally, the 4$\pi$ fractional Josephson effect and their non-Abelian braiding properties in SNS geometries.

\subsubsection*{Signatures in dI/dV}

\begin{figure*}
   \centering
   \includegraphics[width=0.99\textwidth]{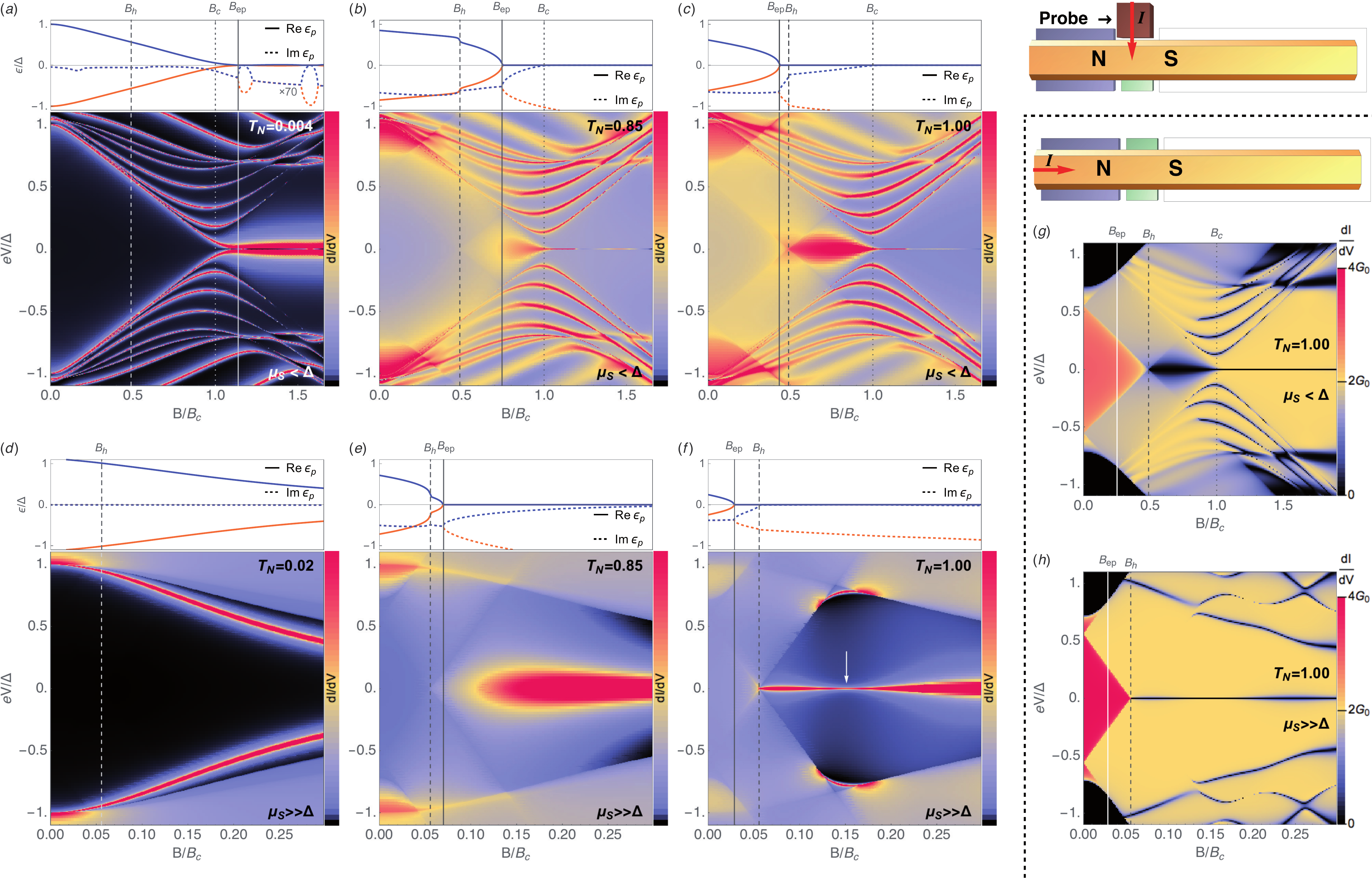}   
   \caption{{\bf Signatures in $dI/dV$}. (a-f) Tunnelling differential conductance $dI/dV$ through a third probe contacted at the junction (brown in Fig. \ref{fig:OpenWire}a), for different junction transparencies $T_N$ and for the same range of Zeeman fields $B$ and bias voltages $V$. First row corresponds to $\mu_S=0.5\Delta$, second row to $\mu_S=10\Delta$, as in Fig. \ref{fig:OpenWire}. The length of the proximitized wire is 1.5 $\mu$m. Evolution of lowest poles across a $B=B_\mathrm{ep}$ exceptional point is shown atop each panel {(higher poles not plotted for clarity)}. Note the sharp zero bias anomaly (ZBA) in the transparent, Andreev limit of panel (f), signalling the presence of MDSs in the topologically trivial regime $B_h<B<B_c$. (g,h) Differential conductance across the junction, for the same parameters as panels (c,f). The ZBA becomes a sharp dip on a constant $2G_0=2e^2/h$ background.}
   \label{fig:dIdV} 
\end{figure*} 

We start by analysing the differential conductance $dI/dV$ through a normal tunnelling probe weakly coupled to the neighbourhood of the junction, brown in Fig. \ref{fig:WireSketch}. In the tunneling limit, this is proportional to the local density of states at the junction at energy $\epsilon=eV$, where $V$ is the bias voltage. (Note that this is different from the differential conductance across the NS contact, which is not necessarily in the tunneling regime, see below). {We computed the tunnelling $dI/dV$ versus $B$ and $V$ using standard quantum transport techniques\cite{Datta:97} (see Appendix). It is} shown in Fig. \ref{fig:dIdV}(a-f) for several transparencies $T_N$, both far from the Andreev limit (top row), and deep into the Andreev limit (bottom row). Atop each panel, the evolution of the lowest complex eigenvalues with $B$ is shown, with an exceptional point bifurcation at $B=B_\mathrm{ep}$. Panel (f), with $T_N\to 1$, corresponds to a cut along the top of Fig. \ref{fig:OpenWire}(e), for which MDSs are fully developed. Their presence gives rise to sharp zero bias anomaly (ZBA) in transport at fields $B>B_h$, with a sharpness that increases exponentially with $\mu_S/\Delta$. This type of ZBA was the first signature of MBSs explored experimentally \cite{Mourik:S12}, though in the present context they arise far from the topological regime $B\ll B_c$. The ZBA is not preceded by signatures of a gap closing. Note also that away from the ideal conditions $T_N\to 1$, $\mu_S\gg \Delta$, wide Majorana resonances are also visible in the topologically trivial regime, panels (b,c,e), albeit of finite lifetime.
Injecting current through the normal reservoir [panels (g,h)] yields a very different $dI/dV$ profile for large $T_N$ (unlike for $T_N\ll 1$), which is no longer a measurement of the local density of states. The ZBA appears in this case as a sharp dip on top of a constant $2e^2/h$ background (perfect Andreev reflection), though again only for $|B|>B_h$. This is consistent with general scattering theory \cite{Beri:PRB09,Fidkowski:PRB12,Pikulin:NJOP12,Pikulin:PRB13,Ioselevich:NJOP13}, that predicts a $dI/dV=0$ from a single mode reservoir at $V\to 0$ if the topology is trivial. The $dI/dV=2e^2/h$ plateau is preceded by another unsplit triangular $4e^2/h$ plateau for $|B|<B_h$, forming a characteristic feature that should be experimentally recognizable. Note that for $B>B_c$ one also obtains a sharp dip at zero, panel (g). This is due to the finite length of the superconducting wire  \cite{Pikulin:JL12,Pikulin:PRB13}, $L_S=1.5 \mu$m here.

\subsubsection*{Spatial localization and Majorana character}

\begin{figure}  
   \centering
   \includegraphics[width=0.98\columnwidth]{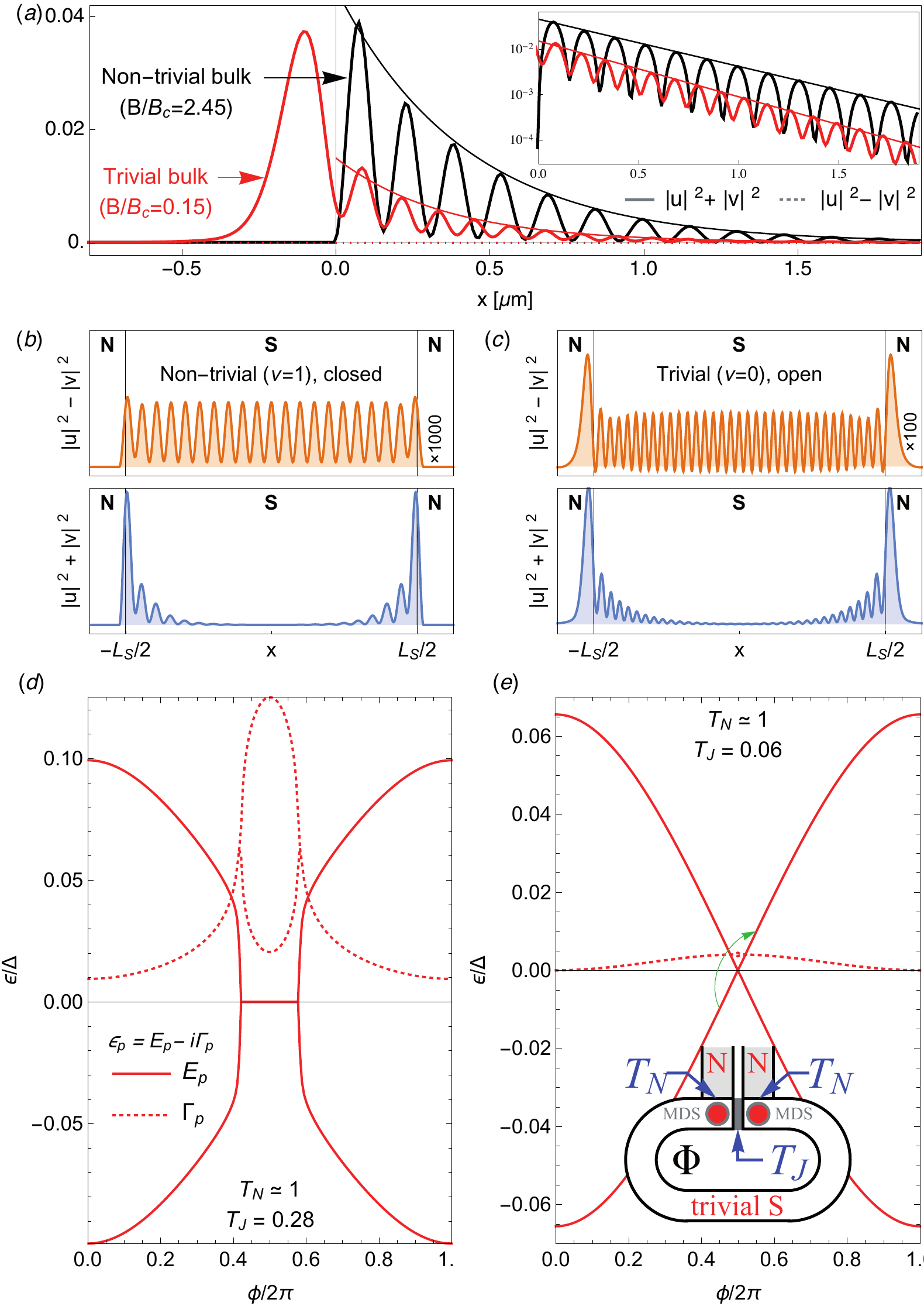}
   \caption{{\bf Spatial localization, Majorana character and 4$\pi$ fractional Josephson effect}. (a) Spatial quasiparticle density $|u|^2+|v|^2$ for Majorana bound states in an NS junction, located at $x=0$. The solid red curve corresponds to an MDS between trivial bulks ($T_N\approx 1$, $B_h<B<B_c$, see white arrow in Fig. \ref{fig:dIdV}f), while the solid black curve corresponds to a conventional MBS in a closed topological wire ($T_N=0$, $B>B_c$). Both decay exponentially $\sim e^{-x/\xi}$ with a Majorana localization length $\xi=\hbar v_F/\Delta(B)$ \cite{Klinovaja:PRB12}  (see inset). Dotted lines are the corresponding charge densities  $|u|^2-|v|^2$, zero everywhere, revealing the Majorana character of both states. (b) Same as (a), albeit in a NSN geometry with finite $L_S$, hosting two overlapping Majoranas. The charge densities $|u|^2-|v|^2$ exhibit spatially uniform oscillations due to the overlap. (d,e) $4\pi$ Josephson effect from MDSs in a Josephson junction of transparency $T_J$, completely open to helical contacts (see inset, $B=B_h<B_c$, $T_N\approx 1$). The MDSs at each contact hybridize across the junction. Their real energies (solid curves) depend on the superconducting phase difference $\phi$, and are zero at $\pi$. The double exceptional point structure around $\phi=\pi$ found here is already a signature in the open regime of an underlying protected crossing in the closed limit (that here would be reached in the $T_N\to 1, T_J\to 0$ limit). While for large $T_J$ (d) the states have a fast decay rate $\Gamma_p$ (dashed curves), the decay is quickly suppressed at small $T_J$ (e). At the same time, the two exceptional points approach and merge at $\phi=\pi$, which then appears as the protected level crossing $E_p=0$ of a closed system. This results in a $4\pi$-Josephson effect like that of topological Josephson junctions, despite the trivial topology.}
   \label{fig:Phenomenology}  
\end{figure}

The spatial locality and Majorana self-conjugation $\gamma=\gamma^\dagger$ are assessed next, by analysing the wavefunction of the MDSs. Figure \ref{fig:Phenomenology}a shows, in red, the quasiparticle density $|\psi(x)|^2=|u|^2+|v|^2$ (solid lines) and charge density $|\rho(x)|^2=|u|^2-|v|^2$ (dashed lines) of the MDS marked by the white arrow in Fig \ref{fig:dIdV}f ($u$ and $v$ are particle and hole components of its wavefunction, respectively). As discussed above, the MDS represents a non-decaying state at zero energy. The figure shows that it is furthermore well localised at the junction, decaying exponentially as $\sim e^{-x/\xi}$ with a Majorana localization length $\xi=\hbar v_F/\Delta(B)$ \cite{Klinovaja:PRB12} into the superconductor (see envelopes and inset in \ref{fig:Phenomenology}a).  For comparison we also show in black the spatial probability $|\psi(x)|^2$ of a conventional $B>B_c$ MBS for a topological bulk at {zero transparency (isolated topological wire)}. For both states, the charge density $\rho(x)$ is zero, as implied by the Majorana relation $\gamma=\gamma^\dagger$. 

We now examine the charge density patterns that arise from the weak overlap of two MDSs. It was shown \cite{Lin:PRB12,Ben-Shach:PRB15} that the charge density $\rho(x)=|u|^2-|v|^2$, which is zero everywhere for an isolated MBS (see Fig. \ref{fig:Phenomenology}a), develops a spatial oscillatory pattern that is uniform throughout space when two MBSs approach each other in a 1D superconductor, irrespective of their particular positions. This is a very specific and non-trivial signature of MBSs that probes the state wavefunction itself, and was proposed as a way to detect MBSs through charge sensing. A transparent NSN junction (with N portions coupled to reservoirs) provides a convenient geometry to study this effect in the topologically trivial regime. Two localized MDSs appear for $|B|>B_h$ at the two ends of the S section. They weakly overlap, and should thus be expected to exhibit uniform spatial charge oscillation throughout the superconductor in analogy to conventional MBSs. Figs. \ref{fig:Phenomenology}(b,c) compare the charge density $\rho(x)$ for $B>B_c$ MBSs in the tunnelling limit and $B_h<B<B_c$ in the transparent limit. Once more, we see the strong similarity between the two cases, which points to an essential equivalence between the two types of states.
 
\subsubsection*{Fractional Josephson effect}
We next consider the fractional Josephson effect. A Josephson junction hosting a pair of conventional MBSs was shown, in the absence of quasiparticle poisoning, to develop an anomalous supercurrent $I(\phi)$ that is $4\pi$-periodic in the superconductor phase difference $\phi$ \cite{Lutchyn:PRL10,Oreg:PRL10,Kitaev:PU01,Fu:PRB09,Law:PRB11,Jiang:PRL11a}. The anomalous supercurrent component is carried by the MBSs, which exhibit a parity-protected zero-energy crossing at $\phi=\pi$.
We consider the analogous Josephson experiment in which a pair of MDSs reside at either side of a topologically trivial Josephson junction.
This may be achieved in an \emph{open} version of a standard Josephson junction formed in a superconducting ring, see inset of Fig. \ref{fig:Phenomenology}e. 
{
The two sides of the weak link, of transmission $T_J$, are coupled to normal reservoirs by two NS contacts that are kept open ($T_N\approx 1$), so that each of them hosts an MDS for $B\gtrsim B_h$.  A flux $\Phi$ through the superconducting loop controls the phase difference $\phi$ across the junction, which sustains a supercurrent $I(\phi)$.} The main contribution to the supercurrent is carried by the $\phi$-dependent hybridization of the two MDSs. As long as $T_J$ is small enough, the MDSs will not decay, since their sibling Majoranas are essentially unperturbed by $T_J$ and remain strongly delocalised into the helical reservoirs. As $T_J$ is increased, however, the MDSs will become a finite-lifetime resonances, as in the case of $T_N<1$.
  
Figure \ref{fig:Phenomenology}d shows the {computed S-matrix poles of the open Josephson junction  from two hybridized decaying Majorana resonances ($T_J$ is large), located at $\epsilon_p$ and $-\epsilon_p^\star$. These poles represent Andreev quasibound states at the junction. Their real energy $\pm E_p$ is shown by the solid lines, and their decay rate by the dashed lines. Each of these poles contribute with a Lorentzian $\frac{1}{\pi}\Gamma_p/[(\omega\mp E_p)^2+\Gamma_p^2]$ to the density of states (DOS) of the junction $\rho(\omega)$. The derivative of the DOS respect to $\phi$, integrated over occupied states ($\omega<0$)} yields the supercurrent carried by these decaying states  \cite{Beenakker:92}, see Appendix. Around $\phi=\pi$, we see that $\pm E_p(\phi)$ crosses two new exceptional points, which in particular force $\pm E_p=0$ at $\phi=\pi$, just like in isolated topological Josephson junctions. In practice, however, the finite decay rate $\Gamma=\mathrm{Im}\,\epsilon_p$ of the hybridized Majorana resonances for large $T_J$ (or smaller $T_N<1$) precludes a fractional Josephson effect from developing, with $4\pi$-periodic harmonics only expected in fast transients. This also happens in finite-length topological nanowires {(see e.g. Andreev spectrum of Fig. \ref{fig:Braiding}a)} and in Josephson junctions with quasiparticle poisoning \cite{Rainis:PRB12}. {We emphasize moreover that the double exceptional point structure around $\phi=\pi$ found here is similarly obtained in the phase dependence of a topologically non-trivial Josephson junction open to an environment \cite{Tarasinski:PRB15}, and has more generally been shown to emerge from Dirac cones in photonic crystals with non-hermicity arising from radiation \cite{Zhen:N15}}. As such it is already a signature in the open regime of an underlying protected crossing in the closed limit (here $T_N\to 1, T_J\to 0$).

{Figure \ref{fig:Phenomenology}e shows a case closer to this limit, with $T_N\approx 1$ and $T_J=0.06$. The suppression of $T_J$ enhances Andreev reflection from the normal leads and suppresses normal reflection processes. The residual lifetime of the hybridized MDSs is similarly suppressed (dashed line). At the same time the two exceptional points approach and merge at $\phi=\pi$, which then appears as the protected level crossing $E_p=0$ of a closed system. Crucially, no number-conserving perturbation can lift this crossing, just split it again into two exceptional points, always with $E_p=0$ at $\phi=\pi$. The two MDSs thus behave as MBSs in a closed topological Josephson junction, although their $\phi=\pi$ crossing is protected by the two underlying exceptional points\cite{Zhen:N15} and charge conjugation symmetry \cite{Malzard:15}, instead of fermion parity (which cannot be defined in an open system). 
We analysed the continuous evolution of the wavefunction through the level crossing at $\phi=\pi$ for $T_J\to 0$}, and found that a system driven slowly through this point (faster than the residual decay rate $\Gamma_0$) will evolve from its ground state to an excited state with 100\% probability (green arrow in panel e). The two hybridized MDSs are indeed \emph{interchanged} upon crossing $\phi=\pi$. As a result, the Josephson current for small $T_J$ becomes $4\pi$-periodic within the residual lifetime $1/\Gamma_0$ of the junction MDSs. Moreover, the critical current scales as $\sqrt{T_J}$. This makes MDS supercurrent indistinguishable from that of conventional MBSs in topological Josephson junctions.

\subsubsection*{Non-Abelian braiding}

\begin{figure} 
   \centering
   \includegraphics[width=0.98\columnwidth]{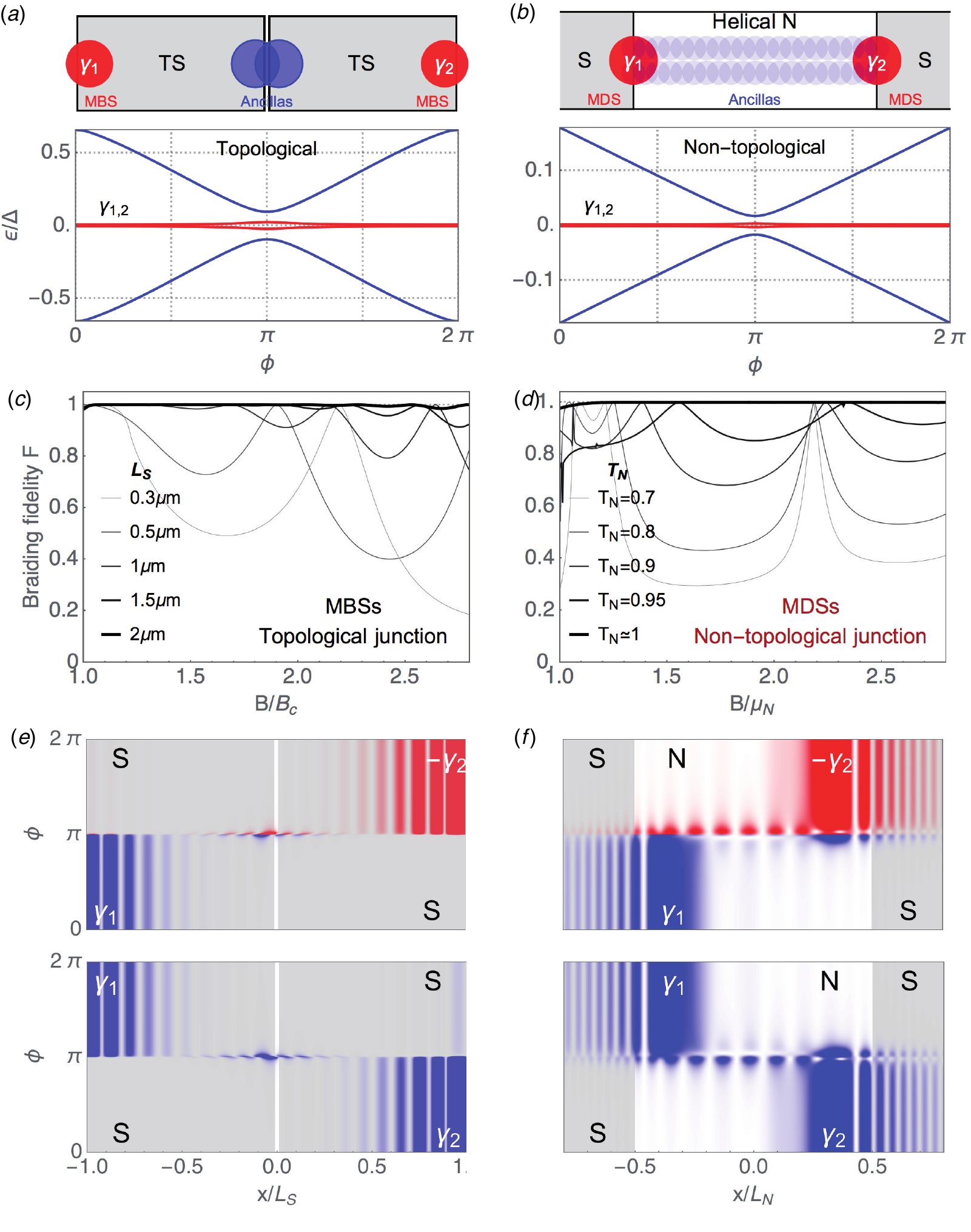}
   \caption{{\bf Braiding of MBSs and MDSs in 1D}. Topological (a) and non-topological (b) four-Majorana Josephson junction (zero energy $\gamma_{1,2}$ Majoranas in red, ancilla Majoranas in blue), and the corresponding low energy Andreev spectrum as a function of superconducting phase difference $\phi$. The $\gamma_{1,2}$ Majoranas undergo braiding as $\phi$ adiabatically increases by $2\pi$. (c,d) Fidelity $F=|\sin 2\phi_B|$ of the $\gamma_{1,2}\to \mp \gamma_{2,1}$ braiding for the two systems in terms of the Berry phase $\pm\phi_B$ of the even/odd ground states upon a $\phi=0\to 2\pi$ adiabatic sweep. The corresponding evolution in real space of $\gamma_{1,2}$ as a function $\phi$, with the relative $\pm 1$ phase respect to the initial basis encoded in blue/red (see main text).}
   \label{fig:Braiding}  
\end{figure}

{We finally analyse the most stringent test of the Majorana character of the MDSs, namely their non-Abelian braiding statistics. This property dictates that upon an adiabatic exchange of two spatially separated (zero energy) MBSs, the corresponding operators $\gamma_{1,2}=\gamma_{1,2}^\dagger$ will undergo a transformation of the type $\gamma_1\to-\gamma_2$ and $\gamma_2\to\gamma_1$. This transformation corresponds to a Berry phase $\pm\pi/4$ for the even/odd degenerate ground states after the adiabatic exchange \cite{Ivanov:PRL01}. If we prepare the system in its even ground state $|e\rangle$ (where `even' refers to the parity of the number of fermions), and perform a braiding operation, it will acquire a $-\pi/4$ global phase, $|e\rangle\to e^{-i\pi/4}|e\rangle$. Starting from the odd ground state, which has an additional zero energy fermion $|o\rangle=(\gamma_1+ i\gamma_2)|e\rangle$, the Berry phase is the opposite $|o\rangle\to e^{i\pi/4}|o\rangle$. This implies that $(\gamma_1+ i\gamma_2)\to e^{2i\phi_B} (\gamma_1+ i\gamma_2)$) under the braiding, with $\phi_B=\pi/4$ irrespective of the details of the exchange path. Together with $\gamma_{1,2}=\gamma_{1,2}^\dagger$, this implies the that $\gamma_1\to-\gamma_2$ and $\gamma_2\to\gamma_1$. More intuitively, if instead of creating a full zero-energy fermion on $|e\rangle$, we just create a Majorana bound state $|\gamma_{1}\rangle=\gamma_{1}|e\rangle$ or $|\gamma_{2}\rangle=\gamma_{2}|e\rangle$ localized at its corresponding boundary, its wavefunction will adiabatically evolve under the exchange process into the \emph{other} Majorana $|\gamma_{1}\rangle\to -|\gamma_{2}\rangle$, and similarly $|\gamma_{2}\rangle\to |\gamma_{1}\rangle$. }

In the following, we demonstrate that MDSs indeed exhibit this exact braiding statistics, just as conventional MBSs. Our aim here is to conceptually demonstrate braiding of EP Majoranas, not to discuss the optimal braiding strategy for practical quantum computing. We have thus taken the simplest possible implementation of braiding, that we explain as follows. While the spatial exchange of two Majoranas in two dimensions is conceptually clear, a subtler, yet equivalent, method is to braid them in parameter space by means of at least one additional MBS pair (ancilla)\cite{Sau:PRB11, Heck:NJOP12,Chiu:EEL15}.  
Remarkably, this idea works even in one dimension \cite{Chiu:EEL15} and braiding of conventional Majoranas can be achieved by sweeping the superconducting phase difference $\phi$ from zero to $2\pi$ in a single Josephson junction between two finite-length topological ($B>B_c$) superconducting wires described by Eq. (\ref{model}), see Fig. \ref{fig:Braiding}a. MBSs at the outer (non-contacted) ends of the wire will form a zero energy fermion for long enough wires (in red), while the inner MBSs (the ancilla pair, in blue) will form a fermion with $\phi$-dependent energy. The adiabatic evolution of the even $|e(\phi)\rangle$ and odd $|o(\phi)\rangle$ ground states as $\phi$ is increased by $2\pi$ indeed results in a Berry phase $e^{\pm i\phi_B}\approx e^{\pm i\pi/4}$ for long $L_S$, leading to exact braiding of the outer Majoranas in this limit. The computed fidelity of the braiding process, defined as $F=|\sin2\phi_B|$, is shown in Fig. \ref{fig:Braiding}c for varying $B>B_c$ and $L_S$. A typical spatial evolution under the braiding process of the two Majoranas {(i.e. the adiabatically evolved $|\gamma(\phi)\rangle$ starting from either $|\gamma_1\rangle$ or $|\gamma_2\rangle$)} is shown in Fig. \ref{fig:Braiding}(e), with blue/red denoting the sign of $\mathrm{Re}(\langle\gamma_1|\gamma(\phi)\rangle+\langle\gamma_2|\gamma(\phi)\rangle)$ that reveals the $|\gamma_{1,2}\rangle\to\mp|\gamma_{2,1}\rangle$ braiding statistics.


The equivalent braiding strategy for MDSs is implemented in a topologically trivial superconductor-helical normal-superconductor Josephson junction \cite{Cayao:PRB15}. The ancilla Majoranas in this situation correspond to the two Majoranas delocalised over the helical region (blue in Fig. \ref{fig:Braiding}b), while the braiding Majoranas are the MDSs localised at each contact (red). Regardless of their different positions in space, the low energy Andreev spectrum in this system is essentially the same as for a four-MBS topological Josephson junction \cite{Cayao:PRB15} (compare Fig. \ref{fig:Braiding}a,b). Computing the adiabatic evolution upon increasing $\phi$ by $2\pi$, we once again find that $\phi_B$ approaches $\pi/4$, i.e. exact braiding statistics like in the topological case, for good enough contact transmission $T_N$. The corresponding braiding fidelity $F$ is shown in Fig. \ref{fig:Braiding}d for increasing $T_N$ and $B$. We see that $F=1$ as $T_N\to 1$. Note the equivalent roles of $T_N$ here and of $L_S$ in the topological case: these are the relevant parameters that control the deviations of the braiding Majoranas from true zero energy bound states in each case, and hence the deviations from exact braiding statistics. The spatial evolution of the braided Majoranas, shown in Fig. \ref{fig:Braiding}f follows a similar pattern as in the topological case, albeit for wavefunctions concentrated at the NS contacts. Hence we conclude that regarding their $|\gamma_{1,2}\rangle\to\mp|\gamma_{2,1}\rangle$ braiding statistics, MDSs behave once more in the same way as conventional MBSs.

\section{Discussion} We have presented a novel approach to engineer Majorana bound states in non-topological superconducting wires. {Instead of inducing a topological transition in a proximitized Rashba wire, we consider a sufficiently transparent normal-superconductor junction created on a Rashba wire, with a topologically trivial superconducting side and a helical normal side. The strong coupling to the helical environment forces a single long-lived resonance to emerge from an exceptional point at precisely zero energy above a threshold transparency. This resonance evolves as the transparency is increased further into a stable dark state localised at the junction.} We have demonstrated this phenomenon both in the multimode Kitaev model and in a realistic model for a proximitized semiconducting nanowire. 

The zero-energy state emerges as the junction traverses an exceptional point at the threshold transparency, and becomes robustly pinned to zero energy without fine tuning by virtue of charge-conjugation symmetry. Moreover, its residual decay rate at perfect transparency is exponentially suppressed in the experimentally relevant Andreev limit.
Finally, we have shown that relevant transport and spectral properties associated to these zero energy states, here dubbed {Majorana dark states}, are indistinguishable from those of conventional Majorana bound states, in particular their braiding statistics. 

{Thus, our proposal offers a new promising strategy towards generating and detecting Majorana bound states in the lab, with potential advantages over more conventional approaches in situations where manipulating the metallic environment and contact properties proves to be simpler than engineering a topological superconducting transition.
Most importantly, the condition for reaching a helical phase in the normal side, while the proximitized region of the nanowire remains in the large $\mu_S$ Andreev limit, is expected to be accessible experimentally. All the necessary ingredients for our proposal are already available in the lab: dramatic advances in fabrication of thin semiconducting nanowires, proximitized with conventional s-wave superconductors, were recently reported \cite{Chang:NN15,Krogstrup:NM15}. Highly transparent, single-channel, NS contacts and a high-quality proximity effect in quantitative agreement with theory were demonstrated. Reaching the helical regime in such high-quality and fully tunable devices should be within reach, so we expect that our proposal for MDSs will be soon tested.}

The general connection demonstrated here between Majorana states and the bifurcation of zero energy complex eigenvalues at exceptional points in open systems offers a new perspective on the mechanisms that may give rise to Majorana states in condensed matter systems. {In this sense it extends conventional strategies based on topological superconductors. It moreover expands on the extensive studies on exceptional point physics in optics, where it has been shown that state coalescence has far-reaching physical consequences, such as e.g. non-Abelian geometric phases \cite{Heiss:EPJD99,Dembowski:PRL01,Kim:FDP13}. To date, most studies of this kind have been concerned with open photonic systems under parity-time (PT) symmetry \cite{Bender:PRL98,Klaiman:PRL08,Regensburger:N12} and with its spontaneous breakdown through exceptional point bifurcations. This leads to intriguing physical phenomena, such as unidirectional transmission or reflection \cite{Regensburger:N12}, loss-induced transparency \cite{Guo:PRL09}, lasers with reversed pump dependence and other exotic properties \cite{Liertzer:PRL12,Brandstetter:NC14,Peng:S14}. Such striking optical phenomena are seemingly unrelated to the physics described in this work, but interesting connections are being made. These include open photonic systems with charge-conjugation symmetry \cite{Malzard:15} (as opposed to PT symmetry), and which show spectral transitions analogous to the zero mode bifurcation discussed here. Also, radiation-induced non-hermicity has been demonstrated as a way to convert Dirac cones into exceptional points \cite{Zhen:N15}, a phenomenon completely analogous to the conversion of zero energy crossings of Bogoliubov-de Gennes excitations in Josephson junctions into double exceptional point structures like those in Fig. \ref{fig:Phenomenology}d. Further research should extend the understanding of these interesting connections.  More importantly, our study and others \cite{Malzard:15,Zhen:N15} raise relevant questions, such as whether there is a deeper connection between topological transitions in closed systems and spectral bifurcations in non-hermitian systems. Answering such questions would help in advancing our understanding of the meaning of non-trivial topology in open systems.}

\section{Acknowledgements}
We are grateful to C. W. J. Beenakker, D. Pikulin, J. Sau and A. Soluyanov for illuminating discussions. We acknowledge the support of the European Research Council and the Spanish Ministry of Economy and Innovation through the JAE-Predoc Program (J. C.), Grants No. FIS2011-23713 (P. S.-J), FIS2012-33521 (R. A.), FIS2013-47328 (E. P.) and the Ramon y Cajal Program (E. P and P. S.-J).

\appendix

\section{Appendix I: Methods}

Transport across the NS junction is computed using the nanowire model for a Rashba wire,
\begin{eqnarray} \label{model}
H_\textrm{S}&=&(2t-\mu_S)\sum_{\sigma n}c_{\sigma n}^\dagger c^{\phantom{\dagger}}_{\sigma n}+\sum_{\sigma n}\Delta\, c_{\sigma n}^\dagger c^\dagger_{\bar\sigma n}+\mathrm{H.c} \\
&&-\sum_{\sigma, \langle n,n'\rangle}t\, c_{\sigma n'}^\dagger c^{\phantom{\dagger}}_{\sigma n} -i\sum_{\sigma,\sigma' \langle n,n'\rangle}t^{\mathrm{SO}\phantom{\dagger}}_{n'-n}c_{\sigma' n'}^\dagger \sigma^{y}_{\sigma'\sigma}c^{\phantom{\dagger}}_{\sigma n}\nonumber\\
&&+\sum_{\sigma,\sigma' n}B\,c_{\sigma' n}^\dagger \sigma^{x}_{\sigma'\sigma}c^{\phantom{\dagger}}_{\sigma n}\nonumber
\end{eqnarray} 
with the non-proximized normal section (N) modelled by the same Hamiltonian, albeit with $\Delta=0$ and a $\mu_N$ in place of $\mu_S$. The normal contact transmission $T^{(n)}_N$ of each incoming mode is computed by also setting  $\Delta=0$  on the proximised (S) side, and using the standard Green's function scheme.  One first splits the system into a left lead (with $\mu_N$), a right lead (with $\mu_S$), and a central section (the interface with a non-uniform profile $\mu(x)$ that transitions from $\mu_N$ into $\mu_S$) coupled to the leads through operators $V_{N/S}$. The total conductance $\mathcal{G}$ of the $M$ incoming modes is then given by Caroli's formula \cite{Caroli:JPCSSP71}
\begin{equation}
\mathcal{G}=\sum_n^M T_N^{(n)}=4\mathcal{G}_0\mathrm{Tr}\left[\Gamma_N G \Gamma_S G^\dagger\right]
\end{equation}
where $\mathcal{G}_0=e^2/h$, $G$ is the dressed retarded Green's function of the central region, $\Gamma_{N/S}=(\Sigma_{N/S}+\Sigma_{N/S}^\dagger)/2$ is the decay operator into the left/right leads, $\Sigma_{N/S}=V_{N/S}^\dagger g_{N/S}V_{N/S}$ is the corresponding self energies, and $g_{N/S}$ is the surface Green's function of the decoupled leads.

{The poles of the scattering matrix presented in the main text are given, close to the origin of the complex plane, by the eigenvalues of non-Hermitian Hamiltonian $H_S+\Sigma(\omega=0)$, where $H_S$ is the (Hermitian) Hamiltonian of a sufficiently long segment of the wire containing the junction, and $\Sigma$ is the self-energy from the remaining wire (the reservoir), that is computed numerically as described above.}

The average normal transmission per mode is defined as $T_N=\mathcal{G}/(M\mathcal{G}_0)$. The values given in the main text were computed for Zeeman $B=0$. $T_N$ depends on the detailed spatial interpolation profile $\mu(x)$ across the interface. An abrupt interface has a smaller transmission than a smooth one, due to the mismatch in Fermi velocity between the two sides. In a real sample, the smoothness of such depletion profile is controlled by geometric parameters of the gating used to deplete the normal side (typically the superconducting side will be difficult to deplete due to screening by the parent superconductor). $T_N$ can also be controlled in a real device by adding a pinch-off gate close to the contact. This possibility is modelled by suppressing a single hopping term $t$ precisely at the contact, where $\Delta(x)$ abruptly jumps from zero to $\Delta$. The combination of mismatch and pinch-off allows to sweep $T_N$ from zero to one.

%

\section{Appendix II: Finite wire effects}
\label{appendix:LN}

\begin{figure} 
   \centering
   \includegraphics[width=0.98\columnwidth]{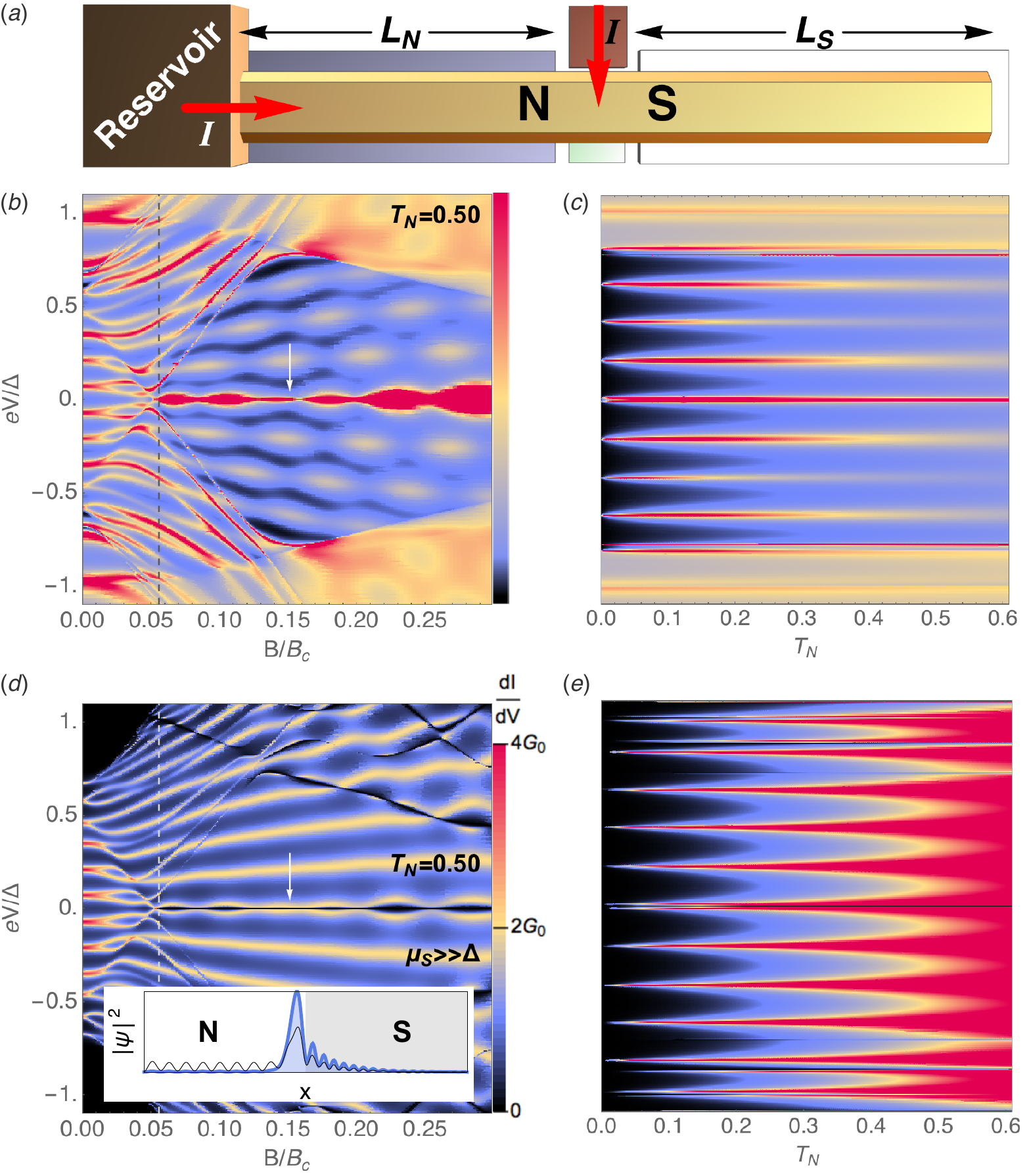}
   \caption{(a) A sketch of a realistic sample with finite wire lengths, $L_N$ and $L_S$. The overall transmission of the junction is a combination of the reservoir-N resistance and the NS resistance. (b) The $dI/dV$ conductance from a third tunneling probe for parameters like in Fig. 4f of the main text, albeit with $L_N=1.5\mu$m and a reduced transmission $T_N=0.5$ coming from the reservoir-N contact resistance. (c) The same $dI/dV$ at a fixed $B$ (white arrow in panel b), versus $T_N$. Unlike the finite bias Andreev resonances, the zero-bias peak from the EP-MBSs is largely insensitive to $T_N$ and $L_N$. (d,e) Differential conductance from the metallic reservoir, instead of the probe (compare to Fig. 4h in the main text). In the inset to (d), spatial profile of the lowest eigenmode of the wire (at the arrow), for $T_N=0.5$ (thick blue line) and $T_N=0$ (thin black line), showing localization of the state at the junction. For the parameters of the simulation, the decay rate of the zero-energy scattering state for $T_N=0.5$ corresponds to $\sim 0.1 \mu eV$.}
   \label{fig:finiteLN}  
\end{figure}

The results presented in the main text for the NS junction in a proximised Rashba wire assumed ideal, infinitely long N and the S sides of the wire. In this section we discuss the corrections that should be expected in real samples with finite lengths $L_S$ and $L_N$, see \ref{fig:finiteLN}a. We show that the emergence of EP-MBSs only weakly depends on these corrections, and is dominated by the properties of the NS contact itself, as described in the main text.

While a finite S length $L_S$ of the proximised wire has no influence in the spectrum of the sample as long as it exceeds its coherence length, the same is not obviously true for $L_N$, particularly if the contact resistance between the N portion of the wire and the macroscopic normal reservoir is large. In such case, the N side will behave like a 1D quantum box between the reservoir and the S side (both modelled with the same $\mu_S\gg\Delta$), rather than as an infinite 1D reservoir for the S side. However, this has little impact on the stabilisation of the EP-MBSs for realistic parameters. In Fig. \ref{fig:finiteLN}b we show the differential conductance $dI/dV$ from a tunnel probe, for the same case as in Fig. 4f in the main text, albeit with a realistic $L_N=1.5\mu$m. The total normal transmission is reduced to $T_N=0.5$ by increasing the reservoir-N contact resistance (the NS junction is still assumed transparent). While considerable structure then arises in the $dI/dV$ due to Fabry-Perot interference effects at finite bias voltage V, the sharp EP-MBS peak (red) is only weakly affected, and remains pinned to zero energy. Plotting the $dI/dV$ versus the reservoir-N transmission, Fig. \ref{fig:finiteLN}c, we see that the EP-MBSs actually remains sharp for any value of $T_N$, unlike the conventional Andreev bound states at finite $V$. This pattern is replicated also for the differential conductance from the reservoir (analogous to Fig. 4h in the main text), shown in Fig. \ref{fig:finiteLN}(d,e). While the finite $L_N$ produces considerable structure at finite bias, the sharp dip at zero bias is insensitive to the value of $L_N$ and $T_N$, as long as the NS interface remains close to transparency.

\section{Appendix III: Interaction effects}
We now briefly discuss our expectations concerning the robustness of our conclusions  against interactions in the normal side. Using renormalization group arguments, Fidkowski et al have demonstrated in Ref. \cite{Fidkowski:PRB12} that the universal low-energy properties of NS junctions (with N described as a Luttinger liquid) are governed by fixed points of either perfect normal reflection or perfect Andreev reflection. In the case of junctions with a trivial superconductor, like the ones discussed here, they demonstrate that perfect Andreev reflection  is unstable for strong negative interactions ($g<2$ in the Luttinger liquid picture). In this case, the low-energy properties of the junction are governed by perfect normal reflection (trivial behaviour) which results in $dI/dV|_{V\rightarrow 0}=0$, as expected. The finite voltage conductance vanishes at small voltages as the power law 
\begin{equation}
G\sim (V/V^*)^{2(2/g-1)},\nonumber
\end{equation}
with $V^*$ a crossover voltage that defines the renormalization group flow from perfect Andreev reflection to perfect normal reflection. This conductance reduction results in a sharp dip in the $dI/dV$, as measured from the helical wire. Note that this dip already occurs without interactions, see Fig. 4h. As we have discussed in the main text, such dip ultimately arises from the formation of a localized dark state, measurable as a quantized $dI/dV$ peak from a third probe, and whose residual lifetime is related to the amplitude of Andreev reflections. This dip has been connected to the so-called Beri degeneracy that predicts a $dI/dV=0$ from a single mode reservoir at $V\to 0$ if the topology is trivial \cite{Beri:PRB09,Fidkowski:PRB12,Pikulin:NJOP12,Pikulin:PRB13,Ioselevich:NJOP13}, and has been shown to appear even in topologically
non-trivial wires of finite length (see Fig. 4g in the main text). The role of interactions, therefore, is to add corrections to that residual lifetime. For an infinitely long helical wire and a perfectly transparent contact, interactions will give a lower bound on the residual decay rate of the dark state.

For a finite-length wire, however, as is relevant for realistic junctions, the Luttinger liquid corrections to the $dI/dV$ are only valid for voltages above $\tilde{V}\sim\hbar v_F/L_N$, the reason being that 
at low enough energies all physical quantities should be more sensitive to the long distance part of the lead (which is a noninteracting Fermi liquid reservoir with $g=1$). Thus, one expects  the conductance to cross over to the noninteracting value  for voltages $V\lesssim\tilde{V}$. This sets a range of voltages where we expect our results to be robust even in the presence of negative interactions in the helical wire. For the parameters of Fig. \ref{fig:finiteLN}(d), $\mu_N=0.14meV$ and $L_N=1.5\mu m$, we estimate $\tilde{V}=\hbar v_F/L_N\sim 25\mu eV=0.1\Delta$ which is much larger than the small residual lifetime around $V\sim 0$. This sets a realistic voltage range $V\lesssim \tilde{V}$ where the renormalization group flow towards normal reflection is cut off. In this voltage range, we expect that our results of EP-MBS in highly transparent trivial junctions should be robust even in the presence of strong negative interactions.

Positive interactions, on the other hand, make the perfect Andreev reflection fixed point stable. Such fixed point stabilised by strong attractions offers an alternative scenario, similar to the one discussed here, where Majorana zero modes may exist in junctions with trivial superconductors \cite{Fidkowski:PRB12}.


\end{document}